# Specification-Driven Benchmarking for Automated Program Repair

## From Static Corpora to Executable Specifications

Yasser Ebrahim[0009-0009-6795-4611]

[1] Algoma University, Brampton ON L6V 1A3, Canada
yasser.ebrahim@algomau.ca

**Abstract.** Automated Program Repair (APR) benchmarks have traditionally been constructed as static datasets whose characteristics are inherited from the defects they contain. While this paradigm has enabled decades of progress, finite corpora provide limited experimental control, become increasingly susceptible to contamination as they are reused, and cannot be systematically regenerated or adapted as evaluation requirements evolve. We propose specification-driven benchmarking, a paradigm in which benchmarks are defined by executable specifications and realized through benchmark generation. The specification explicitly declares the intended properties of the benchmark—including program context, fault taxonomy, difficulty, validation strategy, and corpus constraints—while a generation pipeline realizes those requirements through independent generation, validation, and corpus management components. We develop the conceptual foundations of this approach by introducing a taxonomy of benchmark specification dimensions, establishing how each specification dimension maps to deterministic architectural responsibilities, and arguing that independent validation is a structural requirement for trustworthy benchmark generation. An end-to-end example illustrates how specification choices propagate through the pipeline to produce benchmark instances whose properties are independently verifiable. By treating the benchmark as an executable specification rather than a static dataset, the proposed paradigm shifts benchmark construction from artifact curation to declarative experimental design.

## 1 Introduction

Current APR evaluation relies largely on static benchmark corpora such as Defects4J, BugsInPy, and SWE-bench [1–3]. These benchmarks have enabled reproducible evaluation and comparison of repair systems, but they treat the benchmark as a fixed artifact whose characteristics are determined during dataset construction. Experience with widely used defect datasets also shows that reproducibility and test-suite adequacy can be difficult to preserve in practice [7]. As repair systems continue to improve— particularly with the emergence of large language models—this static paradigm offers limited control over benchmark design and raises concerns about long-term sustainability.

We propose specification-driven benchmark generation, in which researchers define an executable benchmark specification describing the desired properties of repair tasks rather than assembling a fixed collection of defects. A generation pipeline interprets this specification to produce validated benchmark instances on demand. Benchmark characteristics—including fault taxonomy, repair difficulty, validation strategy, and patch constraints—become explicit design parameters rather than inherited properties of a historical dataset. The central premise of this paper is that the benchmark is no longer the dataset, but the executable specification from which benchmark corpora are generated.

This paper introduces the conceptual foundations of specification-driven benchmarking. We define a taxonomy of benchmark specifications, show how each specification dimension governs benchmark generation, explain the role of independent validation, and illustrate the framework through an end-to-end example. s

# 2 Why Static Benchmarks Are Structurally Insufficient

Static benchmark corpora have enabled decades of progress in Automated Program Repair (APR), but their underlying construction paradigm imposes limitations that become increasingly significant for LLM-based repair. These limitations arise not from individual benchmarks or insufficient dataset size, but from representing an effectively unbounded repair space with a finite, publicly distributed corpus. Three structural properties are particularly limiting.

## 2.1 Finite Representation

APR repair tasks vary across many dimensions, including programming language, algorithm family, fault taxonomy, structural complexity, validation strategy, and repair constraints. Any benchmark, regardless of size, samples only a finite subset of this space, with distributions largely determined by the available source projects and historical defects [4, 9]. Increasing benchmark size improves coverage but cannot eliminate this fundamental limitation.

## 2.2 Public Availability

Because LLMs are trained on large collections of public code and repositories, benchmark instances may be encountered during training, making it increasingly difficult to distinguish genuine repair capability from prior exposure. Recent APR- and software-engineering-benchmark work identifies data leakage and contamination as threats to valid evaluation, and reports that fresh task collection can change apparent model performance [5, 6]. Larger benchmarks delay—but do not eliminate—this problem because they remain fixed public artifacts.

### 2.3 Inherited Characteristics

Benchmark properties—including fault distributions, repair difficulty, validation strategy, and structural diversity—are largely inherited from the collected defect corpus rather than explicitly designed [4, 9]. Consequently, researchers cannot independently control these characteristics without constructing a new benchmark. Static benchmarks therefore answer "How does a repair system perform on this corpus?" rather than "How does it perform under the conditions we wish to study?"

These limitations stem from the static benchmark paradigm itself rather than from deficiencies in any individual dataset. This motivates a different perspective: instead of improving static corpora, we make benchmark characteristics explicit through an executable specification that governs benchmark generation.

## 3 The Specification Framework

Specification-driven generation inverts traditional benchmark construction. Instead of collecting defects and characterizing them retrospectively, the researcher first declares the desired benchmark characteristics, and the pipeline generates conforming instances. This declaration—the benchmark specification—is a declarative, machine-executable description of the benchmark. It is not a benchmark itself, but a generative definition: executing the same specification with different random seeds produces distinct yet distributionally equivalent benchmark corpora. Rather than sharing fixed artifacts that may become memorized or contaminated, researchers share an executable description of the intended benchmark.

To serve as the governing artifact for benchmark generation, a specification must capture the essential design decisions of an APR repair task. We identify six orthogonal dimensions describing the software artifact, defect, fault seeding strategy, admissible repairs, correctness criterion, and expected repair difficulty. Together, these dimensions provide a vocabulary for expressing benchmark requirements independently of the underlying generation and validation mechanisms. Table 1 summarizes their role in traditional and specification-driven benchmark construction.

The six dimensions define the semantic characteristics of an APR benchmark.

- **Program context** specifies the execution environment, including programming language, project scope, dependencies, and algorithm family.
- **Fault taxonomy** specifies the semantic defect to be introduced, enabling controlled evaluation across different repair capabilities rather than relying on the fault distribution of historical datasets.
- **Fault seeding** strategy defines how buggy instances are produced, such as LLM synthesis, mutation, or extraction from version histories.
- **Patch constraints** define the admissible repair space through restrictions such as minimal edits, single-hunk patches, or other structural requirements.
- **Validation oracle** specifies the criterion for determining repair correctness during generation and evaluation. Because oracle choice substantially influences benchmark behavior, it is discussed separately in Section 5.

- **Difficulty profile** defines the intended repair complexity using objective structural requirements enforced during validation rather than subjective labels assigned after generation.

**Table 1.** Comparison of benchmark design paradigms for automated program repair.

| Traditional Benchmarking | Specification-Driven Benchmarking |
|---|---|
| Benchmark = Dataset | Benchmark = Executable Specification |
| Characteristics inherited from corpus | Characteristics explicitly declared |
| One published corpus | Unlimited regenerated corpora |
| Fixed after publication | Configurable before generation |

Existing benchmarks instantiate these dimensions implicitly through their collected artifacts. The proposed framework differs by exposing each dimension as an explicit specification parameter that can be independently configured prior to benchmark generation.

# 4 The Specification as Pipeline Controller

The specification's influence is not advisory. Each section of a well-designed spec has a designated, mechanically enforced point in the pipeline. This section makes that mapping explicit -- first through a pipeline overview, then through a structured correspondence table.

## 4.1 Pipeline Architecture

The pipeline comprises five components operating as a closed feedback loop. The Slot Scheduler samples benchmark requirements from the specification. The Prompt Builder translates those requirements into an LLM prompt. The Generator produces candidate instances without enforcing quality constraints. The Validator independently verifies compliance with every specification requirement. Finally, the Corpus Manager enforces corpus-level constraints such as diversity, quotas, and deduplication while routing validation feedback back to the Prompt Builder for informed retries.

Generator and Validator must remain architecturally independent. A component cannot objectively validate its own probabilistic output. The specification provides this separation by defining deterministic acceptance criteria that the Validator enforces independently of the Generator.

## 4.2 Specification-to-Pipeline Mapping

The specification governs benchmark generation by assigning each declared requirement to a unique enforcement point within the pipeline. The Generator attempts to satisfy the specification, while the remaining components independently verify or enforce compliance. Consequently, every specification dimension has an explicit operational role rather than serving as descriptive metadata.

**Table 2.** Mapping specification dimensions to pipeline responsibilities.

| Specification Dimension | Pipeline Responsibility |
|---|---|
| Program Context | Prompt Builder + Corpus Manager |
| Fault Taxonomy | Prompt Builder + Validator |
| Difficulty | Prompt Builder + Validator |
| Quality Constraints | Validator |
| Validation Oracle | Validator |
| Corpus Policies | Slot Scheduler + Corpus Manager |

Program context defines the execution environment and diversity constraints. Fault taxonomy and difficulty guide generation while being independently verified during validation. Quality constraints and the validation oracle determine whether generated instances are admissible, whereas corpus construction policies govern benchmark composition through scheduling, balancing, and duplicate detection.

Without an explicit specification, benchmark characteristics remain implicit and generator-dependent. The specification therefore provides the normative reference against which every generated instance is validated.

### 4.3 Closed-Loop Generation

Specification-driven benchmark generation is inherently iterative. Rather than generating independent repair instances until a target corpus size is reached, each accepted or rejected instance changes the remaining benchmark requirements. Generation therefore operates as a closed-loop process in which the evolving corpus continuously guides subsequent generation.

The Corpus Manager tracks how closely the accepted corpus matches the distributions declared in the specification, including fault subtypes, algorithm families, difficulty tiers, and other corpus-level constraints. Underrepresented regions of the specification space receive higher generation priority, while structured validation feedback is returned to the Prompt Builder so that retries address the deficiencies of previous candidates rather than repeating the same prompt.

This introduces the notion of specification coverage: the degree to which the generated corpus satisfies the declared benchmark specification. Analogous to coverage-guided fuzzing, generation is directed toward unmet regions of the specification space rather than unexplored execution paths [11]. Consequently, the corpus converges toward the benchmark defined by the specification rather than the Generator's natural output distribution.

## 5 Oracle Choice as a Specification Decision

One of the most consequential benchmark design decisions is what constitutes a correct repair. APR evaluation studies have shown that test-suite-based assessment can conflate plausible patches with semantically correct patches [4, 8]. Static benchmarks typically embed this decision within the benchmark artifacts, making the oracle difficult to inspect, modify, or reproduce. Specification-driven generation instead requires the oracle to be declared explicitly as part of the benchmark specification, making it reproducible, auditable, and independent of the generated corpus.

Correctness serves two distinct purposes. During generation, the oracle determines whether a candidate instance is admissible—for example, whether the bug is genuine, the tests are discriminative, and the instance satisfies the specified constraints. During evaluation, it determines whether a repair system's proposed fix is acceptable. These questions need not have identical answers. Encoding both oracles explicitly in the specification prevents ambiguity and applies the same correctness criteria consistently across every benchmark instance.

Different oracle types provide different tradeoffs between rigor and practicality. Test-suite passage is efficient but susceptible to overfitting [8]. Differential testing compares candidate repairs against a reference implementation, allowing multiple behaviorally equivalent fixes [10]. Property checking can offer stronger guarantees through formal invariants, but is applicable only when suitable specifications exist.

Treating oracle choice as part of the benchmark specification provides three advantages. First, correctness criteria become explicit and consistent across the benchmark. Second, the reference implementation serves as a satisfiability witness rather than a unique ground truth, allowing multiple valid repairs when appropriate. Third, the same benchmark specification can be instantiated with different oracle types simply by modifying the specification rather than reconstructing the benchmark, enabling comparative studies that are difficult to perform with static corpora.

## 6 A Concrete End-to-End Example

To demonstrate that the framework is implementable and that specification choices have verifiable consequences, we trace a real instance through the full pipeline. We focus on the result at each stage -- what enters, what is produced, and what the specification contributed. This is a proof of concept, not a claim about production-scale system behavior.

### 6.1 The Specification Slot

The Slot Scheduler draws a combination from the spec's distribution pools.

```
fault_subtype:    logical_operator_error
algorithm_family: accumulation
difficulty_tier:  easy
```

```
language:          Python 3.11  |  granularity: function
validation:        min_tests >= 5, mutation_score >= 80%,
                   coverage>=90%, edit_distance in [1, 10]
```

### 6.2 The Prompt

Prompt excerpt generated by the Prompt Builder, condensed for brevity.

```
User:
Fault: logical_operator_error (boolean op swap in com-
pound condition)
Context: Python 3.11, algorithm_family: accumulation
Difficulty: easy | min_tests>=5 | mutation>=80% | single
hunk
Return: JSON { correct_code, buggy_code, test_suite,
metadata }
```

### 6.3 The Generator's Response

The Generator returns a structured JSON object. The key fields of the produced instance:

```
{ "correct_code": "...score >= min_t and score <=
max_t...",
  "buggy_code":   "...score >= min_t or  score <=
max_t...",  // FAULT
  "fault_subtype": "logical_operator_error", "diffi-
culty_tier": "easy" }
```

### 6.4 Validation

The Validator performs deterministic verification in three sequential tiers. Tier 1 enforces correctness invariants that every instance must satisfy. Tier 2 evaluates configurable quality constraints specified by the benchmark definition. Tier 3 applies heuristic analyses intended to improve semantic quality. Only instances accepted by all tiers are forwarded to the Corpus Manager, which performs duplicate detection and diversity management. This instance passed all checks.

### 6.5 The Stored Entry

The Corpus Manager writes the admitted instance to the output corpus with all validation metrics stored alongside the code artifacts:

```
{ "buggy_code": "...or...", "correct_code": "...and...",
  "metadata": { "subtype": "logical_operator_error",
```

```
"difficulty": "easy", "mutation": 1.0,
"edit_distance": 3, "spec_hash":
"37b592..." } }
```

The spec_hash field links every stored instance to the exact specification that produced it. A researcher holding the spec and this entry can independently reproduce the admission decision without re-invoking the Generator.

### 6.6 BenchCraft: A Concrete Implementation

BenchCraft implements specification-driven benchmark generation for function-level Python APR tasks. Its YAML specification declares program context, fault subtype, difficulty, validation requirements, and corpus constraints. An LLM proposes buggy code, a reference implementation, and a test suite. A separate deterministic Verifier checks test outcomes and configured quality requirements, while a Curator enforces duplicate detection and distribution constraints. The Orchestrator schedules generation slots and returns rejection feedback for subsequent attempts. These components instantiate the architectural responsibilities described in Section 4, turning the specification into enforceable admission and corpus-construction rules.

A preliminary evaluation used Claude Sonnet 4.5 across 12 fault subtypes, six algorithm families, and three structural difficulty tiers. Three runs, each targeting 36 instances with seeds 7, 19, and 88, admitted 33, 35, and 34 instances, respectively. All 102 accepted instances matched their requested fault subtype under the implemented compliance checks. Mean mutation score and line coverage were 95.0% and 98.7%, respectively. The unfilled slots reflect exhausted attempt budgets, illustrating that specification compliance and complete target fulfilment are distinct outcomes.

These results provide initial evidence that an executable specification can govern the construction of validated repair tasks. They do not establish semantic correctness beyond the configured checks or equivalence to real-world defects. Mutation adequacy was measured using three AST-level mutation operators, and difficulty denotes a structural proxy whose relationship to APR-system success has not yet been validated. This brief introduction establishes BenchCraft as an implementation of the proposed concept; the complete system design and empirical evaluation are reserved for a separate paper.

## 7 Implications of Specification-Driven Benchmarking

Specification-driven benchmarking changes the benchmark from a static dataset into an executable specification. This shift has four primary implications:

- **Executable benchmarks.** The benchmark becomes an executable definition of an experiment rather than a finite corpus, enabling unlimited benchmark regeneration while preserving experimental characteristics.
- **Explicit experimental design.** Benchmark properties become declarative design parameters instead of inherited characteristics of historical defect corpora.

- **Adaptive benchmark construction.** Closed-loop generation continuously steers corpus construction toward unmet specification requirements through specification coverage.
- **Specification sharing.** Researchers share executable specifications rather than static datasets, allowing reproducible benchmark generation and controlled experimental variation.

## 8 Related Work

APR evaluation has traditionally relied on curated benchmark corpora such as Defects4J [1], BugsInPy [2], and SWE-bench [3]. Although these benchmarks differ in language, scale, and defect source, they share a common construction paradigm: benchmark instances are collected or synthesized, curated, and released as a fixed dataset. As a result, benchmark characteristics—including fault distributions, repair difficulty, validation strategy, and structural diversity—are largely inherited from the selected defect corpus rather than explicitly specified [4, 9].

Much APR research has focused on improving benchmark quality through larger datasets, broader project coverage, stronger test suites, mutation-based benchmarks, synthetic bug generation, and repository-scale evaluation [4, 9, 12]. These efforts improve benchmark content while preserving the underlying corpus-based methodology. The emergence of LLMs has further exposed limitations of this paradigm, including benchmark contamination, solution leakage, and limited experimental control over benchmark characteristics [5, 6, 14]. Similar challenges have led neighboring areas, including coverage-guided fuzzing and randomized compiler testing, to use generator-driven artifact creation rather than fixed corpora [11, 13].

Our work complements these efforts by addressing how benchmarks are defined rather than which defects they contain. We treat the benchmark as an executable specification that explicitly declares benchmark characteristics before generation, making them controllable design parameters rather than inherited properties of a static corpus. The contribution is therefore a new abstraction for benchmark construction rather than another benchmark corpus.

## 9 Conclusion

This paper proposed specification-driven benchmark generation, a paradigm in which APR benchmarks are defined by executable specifications rather than static datasets. We introduced a taxonomy of benchmark specifications, described how specification dimensions govern generation through deterministic architectural responsibilities, and argued that independent validation is a structural requirement for trustworthy benchmark construction.

By shifting the benchmark from a fixed corpus to an executable specification, benchmark characteristics become explicit, reproducible, and independently controllable. Benchmark corpora become realizations of a specification rather than the benchmark itself.

Realizing this vision will require scalable generation systems and empirical validation, but these challenges do not alter the central conceptual contribution: the benchmark is no longer the dataset, it is the executable specification from which benchmark corpora are generated.